\documentclass{elsart}

\usepackage{graphicx}
\usepackage{amssymb}

\usepackage{cite}

\journal{Physica D: Nonlinear Phenomena}

\begin{document}

\begin{frontmatter}

\centerline{{\Large\sf Physica D 206 (3--4), 252--264 (2005)}}

\bigskip

\hrule \hrule \hrule \hrule \hrule \hrule \hrule \hrule \hrule
\hrule \hrule \hrule \hrule \hrule \hrule \hrule \hrule \hrule
\hrule \hrule \hrule \hrule \hrule \hrule \hrule \hrule \hrule

\bigskip

\title{Time scale synchronization of chaotic oscillators\thanksref{RFBR}}

\thanks[RFBR]{This work is supported by Russian Foundation for Basic Research and U.S.~Civilian
Research \& Development Foundation for the Independent States of
the Former Soviet Union (CRDF), grant {REC--006}.}
\author[SSU,CAS]{Alexander~E.~Hramov}\ead{aeh@cas.ssu.runnet.ru},
\author[SSU,CAS]{Alexey~A.~Koronovskii}\ead{alkor@cas.ssu.runnet.ru}

\address[SSU]{Department of Nonlinear Processes, Saratov State University,
Astrakhanskaya, 83, Saratov, 410012, Russia}
\address[CAS]{College of Applied Science, Saratov State University,
Astrakhanskaya, 83, Saratov, 410012, Russia}

\begin{abstract}
This paper presents the result of the investigation of chaotic
oscillator synchronization. A new approach for detecting of
synchronized behaviour of chaotic oscillators has been proposed.
This approach is based on the analysis of different time scales in
the time series generated by the coupled chaotic oscillators. This
approach has been applied for the coupled R\"ossler and Lorenz
systems.
\end{abstract}

\end{frontmatter}

% main text
\section{Introduction}
\label{Sct:Introduction}

Synchronization of chaotic oscillators is one of the fundamental
phenomena of nonlinear dynamics. It takes place in many
physical~\cite{Parlitz:1996_PhaseSynchroExperimental,%
Tang:1998_PhaseSynchroLasers,Allaria:2001_PhaseSynchroLaser,%
Ticos:2000_PlasmaDischarge,Rosa:2000_PlasmaDischarge} and
biological~\cite{Tass:1998_NeuroSynchro,Anishchenko:2000_humanSynchro}
processes. It seems to play an important role in the ability of
biological oscillators, such as neurons, to act
cooperatively~\cite{Elson:1998_NeronSynchro,Rulkov:2002_2DMap,%
Tass:2003_NeuroSynchro}.

There are several different types of synchronization of coupled
chaotic oscillators which have been described theoretically and
observed experimentally~\cite{Pikovsky:2000_SynchroReview,%
Anishchenko:2002_SynchroEng,Pikovsky:2002_SynhroBook,%
Anshchenko:2001_SynhroBook}. The complete synchronization (CS)
implies coincidence of states of coupled oscillators
$\mathbf{x}_1(t)=\mathbf{x}_2(t)$, the difference between state
vectors of coupled systems converges to zero in the limit
$t\rightarrow\infty$,
\cite{Pecora:1990_ChaosSynchro,%
Pecora:1991_ChaosSynchro,Murali:1994_SynchroIdenticalSyst,%
Murali:1993_SignalTransmission}. It appears if interacting systems
are identical. If the parameters of coupled chaotic oscillators
slightly mismathch, the state vectors are close
$|\mathbf{x}_1(t)-\mathbf{x}_2(t)|\approx 0$, but differ from each
other.

Another type of synchronized behavior of coupled chaotic
oscillators wiht slightly mismatched parameters is the lag
synchronization (LS), when shifted in time the state vectors
coincide with each other, $\mathbf{x}_1(t+\tau)=\mathbf{x}_2(t)$.
When the coupling between oscillator increases the time lag $\tau$
decreases and the synchronization regime tends to be CS
one~\cite{Rosenblum:1997_LagSynchro,Zhigang:2000_GSversusPS,%
Taherion:1999_LagSynchro}.

The generalized synchronization (GS)
\cite{Rulkov:1995_GeneralSynchro,Kocarev:1996_GS,%
Pyragas:1996_WeakAndStrongSynchro}, introduced for drive--responce
systems, means that there is some functional relation between
coupled chaotic oscillators, i.e.
$\mathbf{x}_2(t)=\mathbf{F}[\mathbf{x}_1(t)]$. Finally, it is
necessary to mention the phase synchronization (PS) regime. To
describe the phase synchronization the instantaneous phase
$\phi(t)$ of a chaotic continuous time series is usually
introduced \cite{Pikovsky:2000_SynchroReview,Anishchenko:2002_SynchroEng,%
Pikovsky:2002_SynhroBook,Anshchenko:2001_SynhroBook,%
Rosenblum:1996_PhaseSynchro,Osipov:1997_PhaseSynchro}. The phase
synchronization means the entrainment of phases of chaotic
signals, whereas their amplitudes remain chaotic and uncorrelated.

All synchronization types mentioned above are concerned with each
other (see for
detail~\cite{Parlitz:1996_PhaseSynchroExperimental,%
Rulkov:1995_GeneralSynchro,Zhigang:2000_GSversusPS}), but the
relationship between them is not completely clarified yet. For
each type of synchronization there are its own ways to detect the
synchronized behavior of coupled chaotic oscillators. The complete
synchronization can be displayed by means of comparison of system
state vectors $\mathbf{x}_1(t)$ and $\mathbf{x}_2(t)$, whereas the
lag synchronization can be determined by means of the similarity
function~\cite{Rosenblum:1997_LagSynchro}
\begin{equation}
S^2(\tau)=\frac{\langle|\mathbf{x}_2(t+\tau)-\mathbf{x}_1(t)|^2\rangle}
{\sqrt{\langle|\mathbf{x}_1(t)|^2\rangle\langle|\mathbf{x}_1(t)|^2\rangle}}.
\label{eq:SimilarityFunction}
\end{equation}
If the lag synchronization regime takes place the similarity
function $S(\tau)$ has its minimum $\sigma=\min_\tau S(\tau)=0$
for $\tau$ corresponding to the time shift between the state
vectors\footnote{It is clear, that for the case of the complete
synchronization $S(\tau)$ reaches minimum value $\sigma=0$ for
$\tau=0$.}.

The case of the generalized synchronization is more intricate
because the functional relation $\mathbf{F}[\cdot]$ can be very
complicated, but there are several methods to detect the
synchronized behavior of coupled chaotic oscillators, such as the
auxiliary system approach~\cite{Rulkov:1996_AuxiliarySystem} or
the method of nearest
neighbors~\cite{Rulkov:1995_GeneralSynchro,Pecora:1995_statistics}.

Finally, the phase synchronization of two coupled chaotic
oscillators occurs if the difference between the instantaneous
phases $\phi(t)$ of chaotic signals $\mathbf{x}_{1,2}(t)$ is
bounded by some constant
\begin{equation}
|\phi_{1}(t)-\phi_{2}(t)|<\mathrm{const}.
\label{eq:PhaseLocking}
\end{equation}
It is possible to define a mean frequency
\begin{equation}
\bar{\Omega}=\lim\limits_{t\rightarrow\infty}\frac{\phi(t)}{t}=
\langle\dot{\phi}(t)\rangle, \label{eq:MeanFrequency}
\end{equation}
which should be the same for both coupled chaotic system, i.e.,
the phase locking leads to the frequency entrainment. It is
important to notice, to obtain correct results the mean frequency
$\bar{\Omega}$ of chaotic signal $\mathbf{x}(t)$ should coincide
with the main frequency $\Omega_0=2\pi f_0$ of the Fourier
spectrum (for detail, see~\cite{Anishchenko:2004_ChaosSynchro}).
Unfortunately, there is no general way to introduce the phase for
chaotic time series. There are several approaches which allow to
define the phase for ``good'' systems with simple topology of
chaotic attractor (so--called ``phase coherent attractor''), the
Fourier spectrum of which contains the single main frequency
$f_0$. The example of attractor and Fourier spectrum of such
``good'' system is shown in the figure~\ref{fgr:Rossler}.

First of all, the instantaneous phase $\phi(t)$ can be introduced
as an angle in polar coordinates on the
$(x,y)$--plane~\cite{Pikovsky:1997PhaseSynchro,Rosenblum:1997_LagSynchro}
\begin{equation}
\phi=\arctan\frac{y}{x}, \label{eq:Angle}
\end{equation}
but for that all trajectories of chaotic attractor projection on
the $(x,y)$--plane should revolve around some origin. Sometimes, a
coordinate transformation can be used to obtain a proper
projection~\cite{Pikovsky:1997PhaseSynchro,Pikovsky:2002_SynhroBook}.
Another way to define the phase $\phi(t)$ of chaotic time series
$x(t)$ is the constructing of the analytical
signal~\cite{Pikovsky:2000_SynchroReview,Rosenblum:1996_PhaseSynchro}
\begin{equation}
\zeta(t)=x(t)+j\tilde{x}(t)=A(t)e^{j\phi(t)},
\label{eq:AnalyticSignal}
\end{equation}
where the function $\tilde{x}(t)$ is the Hilbert transform of
$x(t)$
\begin{equation}
\tilde{x}(t)=\frac{1}{\pi}\,\mathrm{P.V.}\int\limits_{-\infty}^{+\infty}\frac{x(\tau)}{t-\tau}\,d\tau
\label{eq:HilbertTransform}
\end{equation}
(where P.V. means that the integral is taken in the sense of the
Cauchy principal value). The instantaneous phase $\phi(t)$ is
defined from~(\ref{eq:AnalyticSignal}) and
(\ref{eq:HilbertTransform}). Moreover, the Poincar\'e secant
surface can be used for the introducing of the instantaneous phase
of chaotic dynamical
system~\cite{Pikovsky:2000_SynchroReview,Rosenblum:1996_PhaseSynchro}
\begin{equation}
\phi(t)=2\pi\frac{t-t_n}{t_{n+1}-t_n}+2\pi n, t_n\leq t \leq
t_{n+1}, \label{eq:PoincareSecant}
\end{equation}
where $t_n$ is the time of the $n$th crossing of the secant
surface by the trajectory. Finally, the phase of chaotic time
series can be introduced by means of the continuous wavelet
transform~\cite{Lachaux:2000_WVTSynchro}, but the appropriate
wavelet function and its parameters should be
chosen~\cite{Quiroga:2002_Kraskov}.

All these approaches give correct results for ``good'' systems
with well--defined phase, but fail for oscillators with
non-revolving trajectories. Such chaotic oscillators are often
called as ``systems with ill--defined phase''. The phase
introducing based on the approaches mentioned above for the system
with ill--defined phase leads usually to incorrect
results~\cite{Anishchenko:2004_ChaosSynchro}. Therefore, the phase
synchronization of such systems can be usually detected by means
of the indirect
indications~\cite{Pikovsky:1997PhaseSynchro,Pikovsky:1996_EurophysLett}
and measurements~\cite{Rosenblum:2002_FrequencyMeasurement}.

In this paper we propose a new approach to detect the
synchronization between two coupled chaotic oscillators. The main
idea of this approach consists in the analysis of the system
behavior on different time scales that allows us to consider
different cases of synchronization from the universal positions.
Using the continuous wavelet transform%
~\cite{Daubechies:1992_WVTBook,Kaiser:1994_Wvt,Torresani:1995_WVT,%
alkor:2003_WVTBookEng} we introduce into consideration the time
scales $s$ and associated with them instantaneous phases
$\phi_s(t)$. As we'll show further, if two chaotic oscillators
demonstrate any type of synchronized behavior, in the time series
$\mathbf{x}_1(t)$ and $\mathbf{x}_2(t)$ generated by these systems
there are necessarily correlated time scales $s$ for which the
phase locking condition
\begin{equation}
|\phi_{s1}(t)-\phi_{s2}(t)|<\mathrm{const}
\label{eq:SPhaseLocking}
\end{equation}
is satisfied.

The structure of this paper is the following. In
section~\ref{Sct:WVTTrans} we discuss the continuous wavelet
transform and the method of the time scales $s$ and associated
with them phases $\phi_s(t)$ definition. In the
section~\ref{Sct:PSSynchro} we consider the case of the phase
synchronization of two mutually coupled R\"ossler systems. We
demonstrate the application of our method and discuss the
relationship between our and traditional approaches. The next
section~\ref{Sct:IllPhase} deals with the synchronization of two
mutually coupled R\"ossler systems with funnel attractors. In this
case the traditional methods for the phase introducing fail and
there is no possibility to detect the phase synchronization
regime, respectively. The synchronization between systems can be
revealed here only by means of the indirect measurements (see for
detail~\cite{Rosenblum:2002_FrequencyMeasurement}). We demonstrate
the efficiency of our method for such cases. In the
section~\ref{Sct:GSSynchro} we consider application of our method
for the unidirectional coupled R\"ossler and Lorenz systems when
the generalized synchronization is observed. In the
section~\ref{Sct:SynchroRelation} we discuss the correlation
between different types of the chaotic synchronization. The final
conclusion is presented in the section~\ref{Sct:Conclusion}.

\section{Continuous wavelet transform}
\label{Sct:WVTTrans}

Let us consider continuous wavelet transform of chaotic time
series $x(t)$
\begin{equation}
W(s,t_0)=\int\limits_{-\infty}^{+\infty}x(t)\psi^*_{s,t_0}(t)\,dt,
\label{eq:WvtTrans}
\end{equation}
where $\psi_{s,t_0}(t)$ is the wavelet--function related to the
mother--wavelet $\psi_{0}(t)$ as
\begin{equation}
\psi_{s,t_0}(t)=\frac{1}{\sqrt{s}}\psi\left(\frac{t-t_0}{s}\right).
\label{eq:Wvt}
\end{equation}
The time scale $s$ corresponds to the width of the wavelet
function $\psi_{s,t_0}(t)$, and $t_0$ is shift of wavelet along
the time axis, the symbol ``$*$'' in~(\ref{eq:WvtTrans}) denotes
complex conjugation. It should be noted that the time scale $s$ is
usually used instead of the frequency $f$ of Fourier
transformation and can be considered as the quantity inversed to
it.

The Morlet--wavelet~\cite{Grossman:1984_Morlet}
\begin{equation}
\psi_0(\eta)=\frac{1}{\sqrt[4]{\pi}}\exp(j\Omega_0\eta)\exp\left(\frac{-\eta^2}{2}\right)
\label{eq:Morlet}
\end{equation}
has been used as a mother--wavelet function. The choice of
parameter value $\Omega_0=2\pi$ provides the relation $s=1/f$
between the time scale $s$ of wavelet transform and frequency $f$
of Fourier transformation.

The wavelet surface
\begin{equation}
W(s,t_0)=|W(s,t_0)|e^{j\phi_s(t_0)} \label{eq:WVT_Phase}
\end{equation}
describes the system's dynamics on every time scale $s$ at the
moment of time $t_0$. The value of $|W(s,t_0)|$ indicates the
presence and intensity of the time scale $s$ mode in the time
series $x(t)$ at the moment of time $t_0$. It is possible to
consider the quantities
\begin{equation}
E(s,t_0)=|W(s,t_0)|^2 \label{eq:Energy}
\end{equation}
and
\begin{equation}
\langle E(s)\rangle=\int|W(s,t_0)|^2\,dt_0, \label{eq:IntEnergy}
\end{equation}
which are instantaneous and integral energy distributions on time
scales, respectively.

At the same time, the phase $\phi_s(t)=\arg\,W(s,t)$ is naturally
introduced for every time scale $s$. It means that it is possible
to describe the behavior of each time scale $s$ by means of its
own phase $\phi_s(t)$. If two interacting chaotic oscillators are
synchronized it means that in time series $\mathbf{x}_1(t)$ and
$\mathbf{x}_2(t)$ there are scales $s$ correlated with each other.
To detect such correlation one can examine the
condition~(\ref{eq:SPhaseLocking}) which should be satisfied for
synchronized time scales.

\section{Phase synchronization of two R\"ossler systems}
\label{Sct:PSSynchro}

Let us start our consideration with two mutually coupled R\"ossler
systems with slightly mismatched
parameters~\cite{Rosenblum:1996_PhaseSynchro,Osipov:1997_PhaseSynchro}
\begin{equation}
\begin{array}{l}
\dot
x_{1,2}=-\omega_{1,2}y_{1,2}-z_{1,2}+\varepsilon(x_{2,1}-x_{1,2}),\\
\dot y_{1,2}=\omega_{1,2}x_{1,2}+ay_{1,2},\\
\dot z_{1,2}=p+z_{1,2}(x_{1,2}-c), \label{eq:Rossler}
\end{array}
\end{equation}
where $a=0.165$, $p=0.2$, and $c=10$. The parameters
$\omega_{1,2}=\omega_0\pm\Delta$ determine the parameter
mistuning, $\varepsilon$ is the coupling parameter.
In~\cite{Rosenblum:1997_LagSynchro} it has been shown that for
these control parameter values and coupling parameter
$\varepsilon=0.05$ the phase synchronization is observed.

\begin{figure}[tbh]
\centerline{\includegraphics*[scale=0.6]{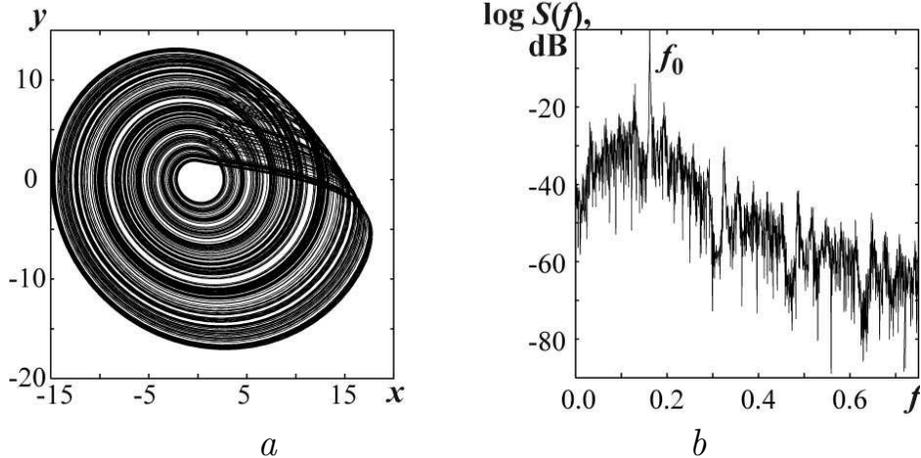}}
\centerline{\large\quad\textit{a}\qquad\qquad\qquad\qquad\qquad\quad\textit{b}}
\caption{(\textit{a}) Phase coherent attractor and (\textit{b})
spectrum of the first R\"ossler system~(\ref{eq:Rossler}).
Coupling parameter $\varepsilon$ between oscillators is equal to
zero \label{fgr:Rossler}}
\end{figure}

For this case the phase of chaotic signal can be easily introduced
in the one of the ways (\ref{eq:Angle})--(\ref{eq:PoincareSecant})
mentioned above, because the phase coherent attractor with rather
simple topological properties is realized in the system phase
space. The attractor projection on the $(x,y)$--plane resembles
the smeared limit cycle where the phase point always rotates
around the origin (Fig.~\ref{fgr:Rossler},\textit{a}). The Fourier
spectrum $S(f)$ contains the main frequency peak $f_0\simeq 0.163$
(see Fig.~\ref{fgr:Rossler},\textit{b}) which coincides with the
mean frequency $\bar{f}=\bar{\Omega}/2\pi$, determinated from the
instantaneous phase $\phi(t)$ dynamics~(\ref{eq:MeanFrequency}).
Therefore, there are no complications to detect the phase
synchronization regime in the two coupled R\"ossler
systems~(\ref{eq:Rossler}) by means of traditional approaches.

When the coupling parameter $\varepsilon$ is equal to $0.05$ the
phase synchronization between chaotic oscillators is observed. The
phase locking condition~(\ref{eq:PhaseLocking}) is satisfied and
the mean frequencies $\bar{\Omega}_{1,2}$ are entrained. So, the
time scales $s_0\simeq 6$ of both chaotic systems corresponding to
the mean frequencies $\bar{\Omega}_{1,2}$ should be correlated
with each other. Correspondingly, the phases $\phi_{s1,2}(t)$
associated with these time scales $s$ should be locked and the
condition~(\ref{eq:SPhaseLocking}) should be satisfied. The time
scales which are the nearest to the time scale $s_0$ should be
also correlated, but the interval of correlated time scales
depends upon the coupling strength. At the same time, should be
time scales which remain uncorrelated. These uncorrelated time
scales cause the difference between chaotic oscillations of
coupled systems.

\begin{figure}[tbh]
\centerline{\includegraphics*[scale=0.6]{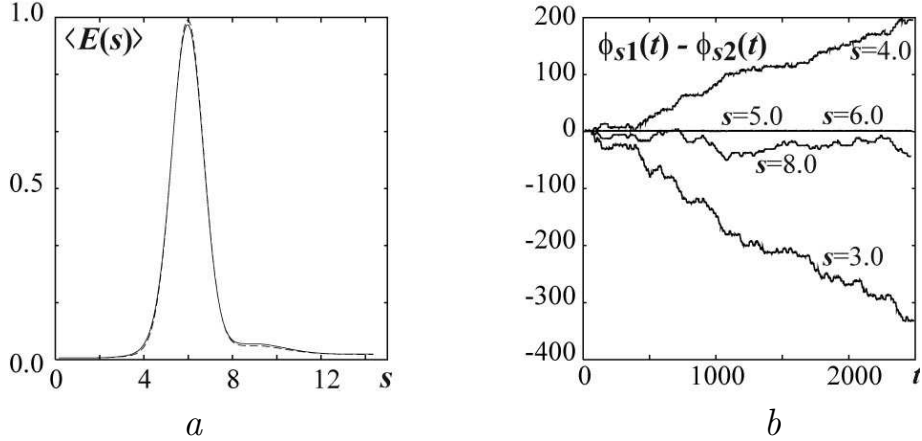}}
\centerline{\large\quad\textit{a}\qquad\qquad\qquad\qquad\qquad\qquad\qquad\quad\textit{b}}
\caption{(\textit{a}) Wavelet power spectrum $\langle E(s)\rangle$
for the first (solid line) and the second (dashed line) R\"ossler
systems~(\ref{eq:Rossler}). (\textit{b}) The dependence of phase
difference $\phi_{s1}(t)-\phi_{s2}(t)$ on time $t$ for different
time scales $s$. The coupling parameter between oscillators is
$\varepsilon=0.05$. The phase synchronization for two coupled
chaotic oscillators is observed \label{fgr:WVTForCoherent}}
\end{figure}

The figure~\ref{fgr:WVTForCoherent} illustrates the behavior of
different time scales for two coupled R\"ossler
systems~(\ref{eq:Rossler}) with phase coherent attractors. It is
clear, that the phase difference $\phi_{s1}(t)-\phi_{s2}(t)$ for
scales $s_0=6$ is bounded and, therefore, time scales $s_0=6$
corresponding to the main frequency of Fourier spectrum $f_0$ are
synchronized. It is important to note that wavelet power spectra
$\langle E_{1,2}(s)\rangle$ are close to each other (see
Fig.~\ref{fgr:WVTForCoherent},\textit{a}) and time scales $s$
characterized by the large value of energy (e.g., s=5) which are
close to the main time scale $s_0=6.0$ are correlated, too. There
are also time scales which aren't synchronized, like $s=3.0$,
$s=4.0$, etc. (see Fig.~\ref{fgr:WVTForCoherent},\textit{b}).

So, when two mutually coupled chaotic oscillators with phase
coherent attractors are considered, the traditional method based
on the instantaneous phase $\phi(t)$ of chaotic signal introducing
and our approach lead to the equivalent results.

\section{Synchronization of two R\"ossler systems with funnel attractors}
\label{Sct:IllPhase}

Let us consider more complicated example when it is impossible to
correctly introduce the instantaneous phase $\phi(t)$ of chaotic
signal $\mathbf{x}(t)$. It is clear, that for such cases the
traditional methods of the phase synchronization detecting fail
and it is necessary to use another techniques, e.g., like indirect
measurements~\cite{Rosenblum:2002_FrequencyMeasurement}. On the
contrary, our approach gives correct results and allows to detect
the synchronization between chaotic oscillators easily as before.

\begin{figure}[bh]
\centerline{\includegraphics*[scale=0.6]{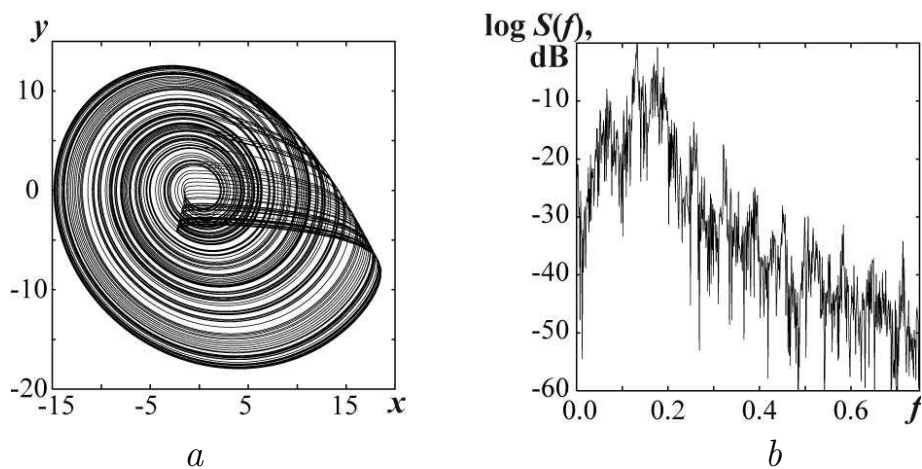}}
\centerline{\large\quad\textit{a}\qquad\qquad\qquad\qquad\qquad\qquad\qquad\quad\textit{b}}
\caption{(\textit{a}) Phase picture and (\textit{b}) power
spectrum of the first R\"ossler system~(\ref{eq:FunnelRoessler})
oscillations. Coupling parameter $\varepsilon$ is equal to zero
\label{fgr:FunnelRoessler}}
\end{figure}

To illustrate it we consider two non--identical coupled R\"ossler
systems with funnel attractors (Fig.~\ref{fgr:FunnelRoessler}):
\begin{equation}
\begin{array}{l}
\dot
x_{1,2}=-\omega_{1,2}y_{1,2}-z_{1,2}+\varepsilon(x_{2,1}-x_{1,2}),\\
\dot y_{1,2}=\omega_{1,2}x_{1,2}+ay_{1,2}+\varepsilon(y_{2,1}-y_{1,2}),\\
\dot z_{1,2}=p+z_{1,2}(x_{1,2}-c), \label{eq:FunnelRoessler}
\end{array}
\end{equation}
where $\varepsilon$ is a coupling parameter, $\omega_1=0.98$,
$\omega_2=1.03$. The control parameter values have been selected
by analogy with~\cite{Rosenblum:2002_FrequencyMeasurement} as
$a=0.22$, $p=0.1$, $c=8.5$.

\begin{figure}[tbh]
\centerline{\includegraphics*[scale=0.6]{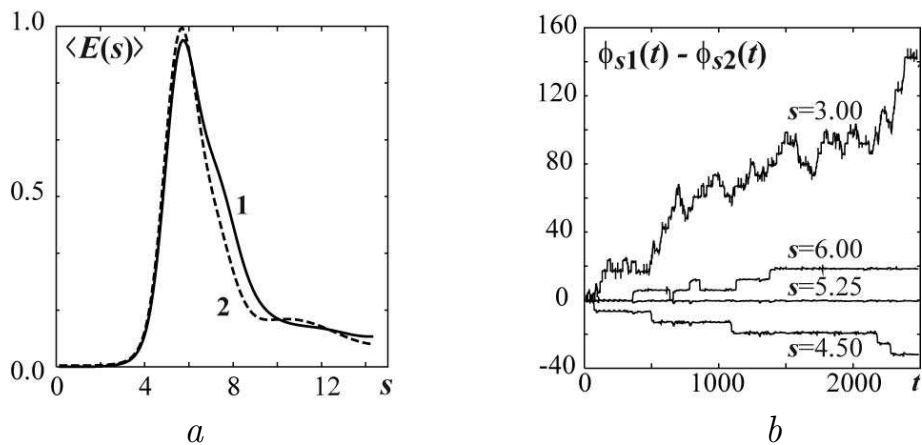}}
\centerline{\large\quad\textit{a}\qquad\qquad\qquad\qquad\qquad\qquad\qquad\quad\textit{b}}
\caption{(\textit{a}) The normalized energy distribution in
wavelet spectrum $\langle E(s)\rangle$ for the first (the solid
line denoted as ``1'') and the second (the dashed line denoted as
``2'') R\"ossler systems~(\ref{eq:FunnelRoessler}); (\textit{b})
the phase difference $\phi_{s1}(t)-\phi_{s2}(t)$ for two coupled
R\"ossler systems. The value of coupling parameter has been
selected as $\varepsilon=0.05$. The time scales $s=5.25$ are
correlated with each other and the synchronization has been
observed \label{fgr:FunRossK=0_05}}
\end{figure}

In~\cite{Rosenblum:2002_FrequencyMeasurement} it has been shown by
means of the indirect measurements that for the coupling parameter
value $\varepsilon=0.05$ the synchronization of two mutually
coupled R\"ossler systems~(\ref{eq:FunnelRoessler}) takes place.
Our approach based on the analysis of the dynamics of different
time scales $s$ gives analogous results. So, the behavior of the
phase difference $\phi_{s1}(t)-\phi_{s2}(t)$ for this case has
been presented in the figure~\ref{fgr:FunRossK=0_05},\textit{b}.
One can see that the phase locking takes place for the time scales
$s=5.25$ which are characterized by the largest energy value in
the wavelet power spectra $\langle E(s)\rangle$ (see
Fig.~\ref{fgr:FunRossK=0_05},\textit{a}). Thus, we can say that
the time scales $s=5.25$ of two oscillators are synchronized with
each other. It is important to note that the other time scales
(e.g., $s=4.5, 6.0$ et. al.) remain uncorrelated. For such time
scales the phase locking has not been observed (see
Fig.~\ref{fgr:FunRossK=0_05},\textit{b}).

It is clear, that the mechanism of the synchronization of coupled
chaotic oscillators is the same in both cases considered in the
sections~\ref{Sct:PSSynchro} and \ref{Sct:IllPhase}. The
synchronization phenomenon is caused by the existence of time
scales $s$ in system dynamics correlated with each other.
Therefore, there is no reason to divide considered synchronization
examples into different types.

\section{Generalized synchronization of R\"ossler and Lorenz systems}
\label{Sct:GSSynchro}

Let us consider another type of synchronized behavior, so--called
the generalized synchronization. As oscillator samples the
R\"ossler (drive) and Lorenz (response) systems have been
selected. The equations of the drive system
$\mathbf{x}_1=(x_1,y_1,z_1)^T$ are
\begin{equation}
\begin{array}{l}
\dot x_{1}=-\omega y_{1}-z_{1},\\
\dot y_{1}=\omega x_{1}+ay_{1},\\
\dot z_{1}=p+z_{1}(x_{1}-c), \label{eq:GSRossler}
\end{array}
\end{equation}
and the response system $\mathbf{x}_2=(x_2,y_2,z_2)^T$ is given by
\begin{equation}
\begin{array}{l}
\dot x_{2}=-\sigma(x_{2}-y_{2}),\\
\dot y_{2}=ru(t)-y_2-u(t)z_2,\\
\dot z_{2}=u(t)y_2-bz_2. \label{eq:GSLorenz}
\end{array}
\end{equation}
The drive signal $u(t)$ and control parameter values have been
selected by analogy with~\cite{Kocarev:1996_GS} as
$u(t)=x_1+y_1+z_1$, $p=2$, $c=4$, $\omega=1$, $a=1$, $\sigma=10$,
$r=28$, $b=2.666$. It has been analytically shown
(see~\cite{Kocarev:1996_GS}) that the response
system~(\ref{eq:GSLorenz}) is asymptotical stable for arbitrary
drive signals $u(t)$ and arbitrary initial conditions, and,
therefore the generalized synchronization always occurs although
drive and response systems are completely different. Obviously,
the generalized synchronization regime can be also detected by
means of other numerical or experimental methods (e.g., the
axillary system approach).

\begin{figure}[tbh]
\centerline{\includegraphics*[scale=0.6]{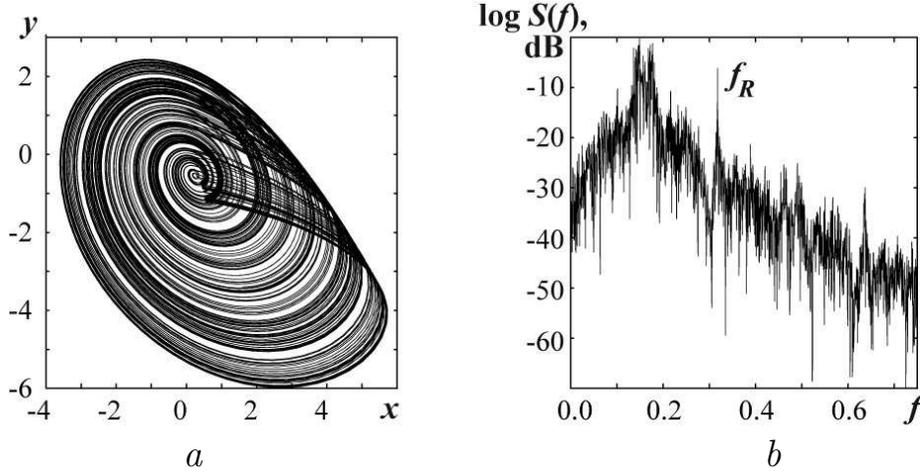}}
\centerline{\large\quad\textit{a}\qquad\qquad\qquad\qquad\qquad\qquad\qquad\quad\textit{b}}
\caption{(\textit{a}) Chaotic attractor and (\textit{b}) spectrum
of the R\"ossler (drive) system~(\ref{eq:GSRossler}).
\label{fgr:GSRossler}}
\end{figure}

\begin{figure}[tbh]
\centerline{\includegraphics*[scale=0.6]{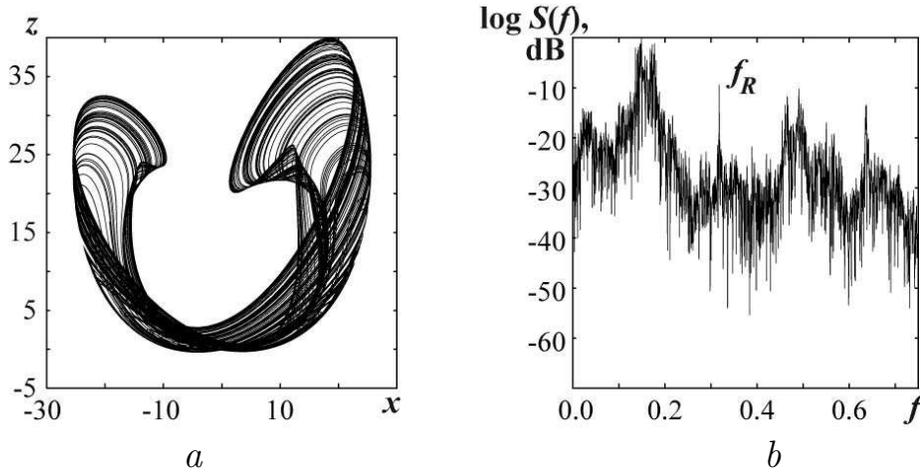}}
\centerline{\large\quad\textit{a}\qquad\qquad\qquad\qquad\qquad\qquad\qquad\quad\textit{b}}
\caption{(\textit{a}) Chaotic attractor and (\textit{b}) spectrum
of the Lorenz (response) system~(\ref{eq:GSLorenz}).
\label{fgr:GSLorenz}}
\end{figure}

The chaotic attractors and corresponding them power spectra $S(f)$
of unidirectional coupled R\"ossler and Lorenz systems are shown
in the figures~\ref{fgr:GSRossler} and \ref{fgr:GSLorenz},
respectively. One can see in the Fourier spectrum of the Lorenz
system the presence of the peaks corresponding to the frequencies
of the R\"ossler system oscillations
(see~Fig.~\ref{fgr:GSRossler},\textit{b} and
\ref{fgr:GSLorenz},\textit{b}). Therefore, the time scales $s$ of
coupled systems can be correlated with each other if chaotic
oscillators are synchronized on these time scales.

\begin{figure}[tbh]
\centerline{\includegraphics*[scale=0.6]{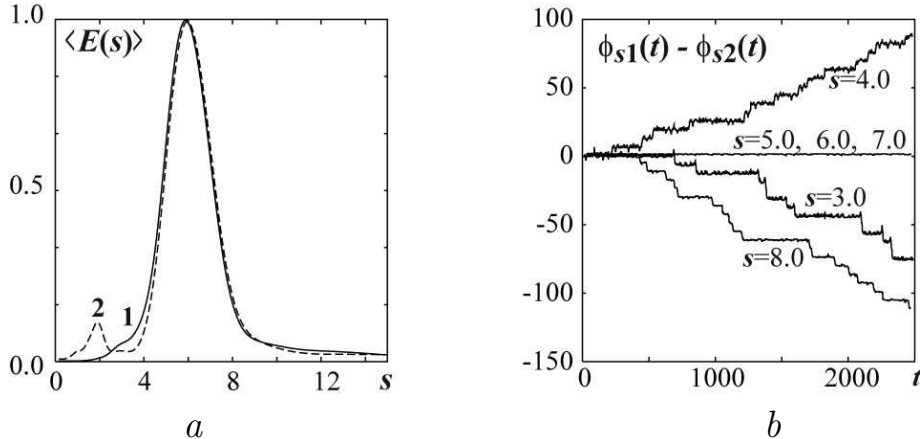}}
\centerline{\large\quad\textit{a}\qquad\qquad\qquad\qquad\qquad\qquad\qquad\quad\textit{b}}
\caption{(\textit{a}) The normalized energy distribution in
wavelet spectrum $\langle E(s)\rangle$ for the
R\"ossler~(\ref{eq:GSRossler}) (the solid line denoted as ``1'')
and Lorenz~(\ref{eq:GSLorenz}) (the dashed line denoted as ``2'')
systems. Each energy distribution has been normalized on its own
coefficient; (\textit{b}) the phase difference
$\phi_{s1}(t)-\phi_{s2}(t)$ for coupled systems, when the
generalized synchronization regime takes place. The time scales
$s=5$, $s=6$, $s=7$ are correlated with each other whereas the
other time scales (e.g., $s=3$, $s=8$, etc.) aren't synchronized
\label{fgr:GSSynchro}}
\end{figure}

The dependencies of phase difference $\phi_{s1}(t)-\phi_{s2}(t)$
on time for different time scales $s$ are shown in the
figure~\ref{fgr:GSSynchro}. It is clear, that there is the range
of scales (approximately $s=5\div 7$) the behavior of which is
synchronized. At the same time there are time scales which are not
synchronized, as in the case of the phase synchronization (see.
e.g., Fig.~\ref{fgr:WVTForCoherent}). It is important to note,
despite the fact that there is the frequency $f_R\simeq 0.33$ in
the spectra of both R\"ossler and Lorenz systems (see
Fig.~\ref{fgr:GSRossler},\textit{b} and
\ref{fgr:GSLorenz},\textit{b}), the time scales $s_R=1/f_R\simeq
3$ of coupled oscillators are not synchronized
(Fig.~\ref{fgr:GSSynchro},\textit{b}). So, the presence of
equivalent frequencies $f$ in the power spectra of interacting
systems doesn't mean the obligatory synchronization on the time
scale $s_f=1/f$ corresponding to this frequency.

Thus, the generalized synchronization of two unidirectionally
coupled absolutely different system is revealed as the synchronous
dynamics of several time scales on which the phases $\phi_s(t)$
are locked. One can see that different types of the
synchronization such as PS and GS are quite similar to each other
when the time scale behavior of interacting systems is analysed.
So, the time scales of R\"ossler and Lorenz systems in the
generalized synchronization regime behave themselves equivalently
to the cases of the phase synchronization of two R\"ossler
systems~(\ref{eq:Rossler}), although for the considered
generalized synchronization case we can not correctly introduce
the instantaneous phase $\phi(t)$ for neither R\"ossler nor Lorenz
systems (see Fig.~\ref{fgr:GSRossler} and \ref{fgr:GSLorenz}) at
all. Obviously, it is necessary to consider the correlation
between different types of synchronization and transitions from
one of them to another one. This topic will be discussed in the
next section.

\section{From unsynchronized behavior to complete synchronization regime}
\label{Sct:SynchroRelation}

It has been shown in~\cite{Rosenblum:1997_LagSynchro} that there
is certain relationship between PS, LS and CS for chaotic
oscillators with slightly mismatched parameters. With the increase
of coupling strength the systems undergo the transition from
unsynchronized chaotic oscillations to the phase synchronization.
With a further increase of coupling the lag synchronization is
observed. The next increasing of the coupling parameter leads to
the decreasing of the time lag and both systems tend to have the
complete synchronization regime.

\begin{figure}[tbh]
\centerline{\includegraphics*[scale=0.6]{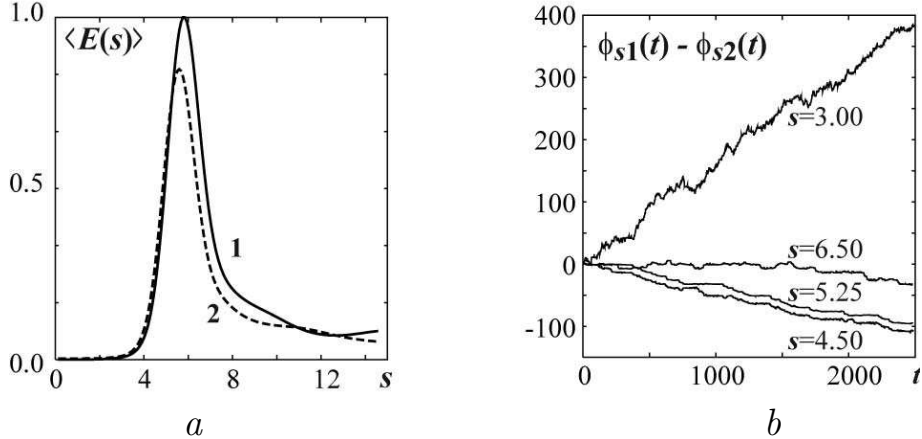}}
\centerline{\large\quad\textit{a}\qquad\qquad\qquad\qquad\qquad\qquad\qquad\quad\textit{b}}
\caption{(\textit{a}) The normalized energy distribution in
wavelet spectrum $\langle E(s)\rangle$ for the first (the solid
line denoted as ``1'') and the second (the dashed line denoted as
``2'') R\"ossler systems; (\textit{b}) the phase difference
$\phi_{s1}(t)-\phi_{s2}(t)$ for two coupled R\"ossler systems. The
value of coupling parameter has been selected as
$\varepsilon=0.025$. There is no phase synchronization between
systems \label{fgr:FunRossK=0_025}}
\end{figure}

Let us consider the dynamics of different time scales $s$ of two
nonidentical mutually coupled R\"ossler
systems~(\ref{eq:FunnelRoessler}). If there is no phase
synchronization between the oscillators, then their dynamics
remain uncorrelated for all time scales $s$. The
figure~\ref{fgr:FunRossK=0_025} illustrates the dynamics of two
coupled R\"ossler systems when the coupling parameter
$\varepsilon$ is enough small ($\varepsilon=0.025$). The power
spectra $\langle E(s)\rangle$ of wavelet transform for R\"ossler
systems differ from each other
(Fig.~\ref{fgr:FunRossK=0_025},\textit{a}), but the maximum values
of the energy correspond approximately to the same time scale $s$
in both systems. It is clear, that the phase difference
$\phi_{s1}(t)-\phi_{s2}(t)$ is not bounded for all time scales
(see Fig.~\ref{fgr:FunRossK=0_025},\textit{b}). It means that
there are no time scales $s$ correlated with each other and two
R\"ossler systems aren't synchronized at all.

As soon as any of the time scales of the first chaotic oscillator
becomes correlated with the other one of the second oscillator
(e.g., when the coupling parameter increases), the phase
synchronization occurs (see Fig.~\ref{fgr:FunRossK=0_05} in the
section~\ref{Sct:IllPhase}). It is clear, that the time scales $s$
characterized by the largest value of energy in wavelet spectrum
$\langle E(s)\rangle$ become correlated first. The other time
scales remain uncorrelated as before. The phase synchronization
between chaotic oscillators leads to the phase
locking~(\ref{eq:SPhaseLocking}) on the correlated time scales
$s$.

\begin{figure}[tbh]
\centerline{\includegraphics*[scale=0.6]{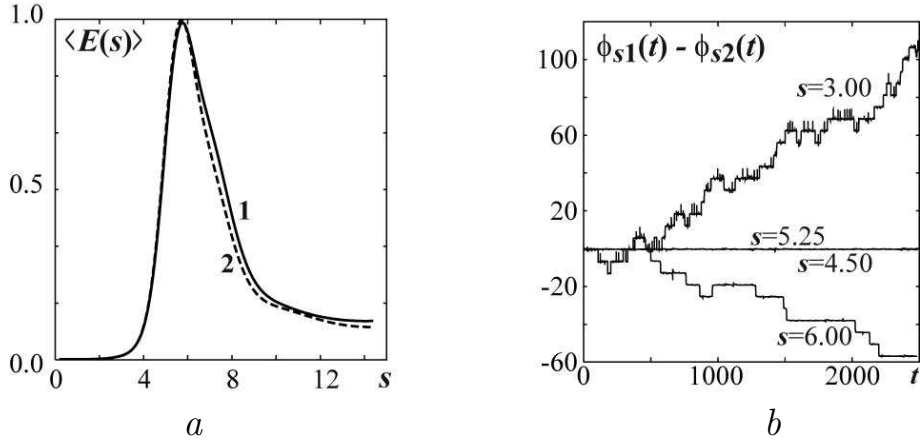}}
\centerline{\large\quad\textit{a}\qquad\qquad\qquad\qquad\qquad\qquad\qquad\quad\textit{b}}
\caption{(\textit{a}) The normalized energy distribution in
wavelet spectrum $\langle E(s)\rangle$ for the first (the solid
line denoted as ``1'') and the second (the dashed line denoted as
``2'') R\"ossler systems; (\textit{b}) the phase difference
$\phi_{s1}(t)-\phi_{s2}(t)$ for two coupled R\"ossler systems. The
value of coupling parameter has been selected as
$\varepsilon=0.07$. \label{fgr:FunRossK=0_07}}
\end{figure}

When the parameter of coupling between chaotic oscillators
increases, more and more time scales become correlated and one can
say that the degree of the synchronization grows. So, with the
further increasing of the coupling parameter value (e.g.,
$\varepsilon=0.07$) in the coupled R\"ossler
systems~(\ref{eq:FunnelRoessler}) the time scales which were
uncorrelated before become synchronized (see
Fig.~\ref{fgr:FunRossK=0_07},\textit{b}). One can see that the
time scales $s=4.5$ are synchronized in comparison with the
previous case ($\varepsilon=0.05$,
Fig.~\ref{fgr:FunRossK=0_05},\textit{b}) when these time scales
were uncorrelated. The number of time scales $s$ demonstrating the
phase locking increases, but there are nonsynchronized time scales
as before (e.g., the time scales $s=3.0$ and $s=6.0$ remain still
nonsynchronized).

\begin{figure}[tbh]
\centerline{\includegraphics*[scale=0.6]{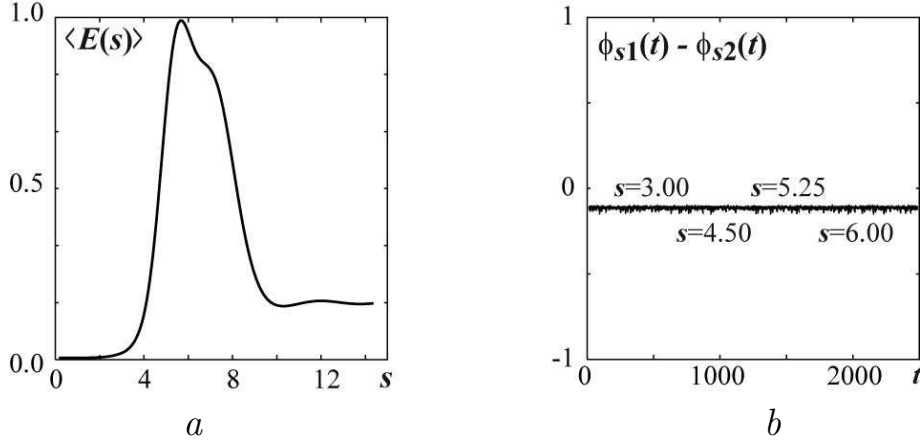}}
\centerline{\large\quad\textit{a}\qquad\qquad\qquad\qquad\qquad\qquad\qquad\quad\textit{b}}
\caption{(\textit{a}) The normalized energy distribution in
wavelet spectrum $\langle E(s)\rangle$ for the R\"ossler system;
(\textit{b}) the phase difference $\phi_{s1}(t)-\phi_{s2}(t)$ for
two coupled R\"ossler systems. The value of coupling parameter has
been selected as $\varepsilon=0.25$. The lag synchronization has
been observed, all time scales are synchronized
\label{fgr:FunRossK=0_25}}
\end{figure}

The arising of the lag
synchronization~\cite{Rosenblum:1997_LagSynchro} between
oscillators means that all time scales are correlated. Indeed,
from the condition of the lag--synchronization ${x_1(t-\tau)\simeq
x_2(t)}$ one can obtain that ${W_1(s,t-\tau)\simeq W_2(t,s)}$ and,
therefore, ${\phi_{s1}(t-\tau)\simeq\phi_{s2}(t)}$. It is clear,
in this case the phase locking condition~(\ref{eq:PhaseLocking})
is satisfied for all time scales $s$. E.g., when the coupling
parameter of chaotic oscillators~(\ref{eq:FunnelRoessler}) becomes
enough large ($s=0.25$) the lag synchronization of two coupled
oscillators appears. In this case the power spectra of wavelet
transform coincide with each other (see
Fig.~\ref{fgr:FunRossK=0_25},\textit{a}) and the phase locking
takes place for all time scale $s$
(Fig.~\ref{fgr:FunRossK=0_25},\textit{b}). It is important to note
that the phase difference $\phi_{s1}(t)-\phi_{s2}(t)$ is not equal
to zero for the case of the lag synchronization. It is clear that
this difference depends on the time lag $\tau$.

The next increasing of the coupling parameter leads to the
decreasing of the time lag
$\tau$~\cite{Rosenblum:1997_LagSynchro}. Both systems tend to have
the complete synchronization regime $x_1(t)\simeq x_2(t)$,
therefore the phase difference $\phi_{s1}(t)-\phi_{s2}(t)$ tends
to be a zero for all time scales.

So, it is clear, that PS, LS and CS are naturally interrelated
with each other and the synchronization type depends on the number
of synchronized time scales. At the same time, the relationship
between PS and GS is not clear in detail. There are several
works~\cite{Parlitz:1996_PhaseSynchroExperimental,%
Zhigang:2000_GSversusPS} dealing with the problem, how GS and PS
are correlated with each other. E.g.,
in~\cite{Zhigang:2000_GSversusPS} it has been reported that two
unidirectional coupled R\"ossler systems can demonstrate the
generalized synchronization while the phase synchronization has
not been observed. This case allows to be easily explained by
means of the time scale analysis. The equations of R\"ossler
system are
\begin{equation}
\begin{array}{l}
\dot x_{1}=-\omega_{1}y_{1}-z_{1},\\
\dot y_{1}=\omega_{1}x_{1}+ay_{1},\\
\dot z_{1}=p+z_{1}(x_{1}-c)\\
\dot x_{2}=-\omega_{2}y_{2}-z_{2}+\varepsilon(x_1-x_2),\\
\dot y_{2}=\omega_{2}x_{2}+ay_{2},\\
\dot z_{2}=p+z_{2}(x_{2}-c),\\
\label{eq:UniDirecRosslers}
\end{array}
\end{equation}
where $\mathbf{x}_1=(x_1,y_1,z_1)^T$ and
$\mathbf{x}_2=(x_2,y_2,z_2)^T$ are the state vectors of the first
(drive) and the second (response) R\"ossler systems, respectively.
The control parameter values have been chosen as $\omega_1=0.8$,
$\omega_2=1.0$, $a=0.15$, $p=0.2$, $c=10$ and $\varepsilon=0.2$.
The generalized synchronization takes place in this case
(see~\cite{Zhigang:2000_GSversusPS} for detail). Why it is
impossible to detect the phase synchronization in the
system~(\ref{eq:UniDirecRosslers}) although the generalized
synchronization is observed becomes clear from the time scale
analysis.

\begin{figure}[tbh]
\centerline{\includegraphics*[scale=0.6]{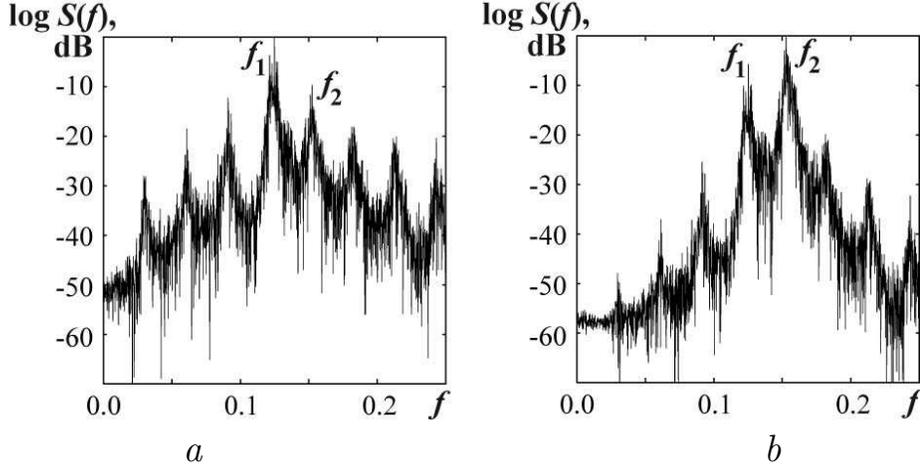}}
\centerline{\large\quad\textit{a}\qquad\qquad\qquad\qquad\qquad\qquad\qquad\quad\textit{b}}
\caption{Fourier spectra for(\textit{a}) the first (drive) and
(\textit{b}) the second (response) R\"osler
systems~(\ref{eq:UniDirecRosslers}). The coupling parameter is
$\varepsilon=0.2$. The generalized synchronization takes place
\label{fgr:GS2_spectra}}
\end{figure}

\begin{figure}[tbh]
\centerline{\includegraphics*[scale=0.6]{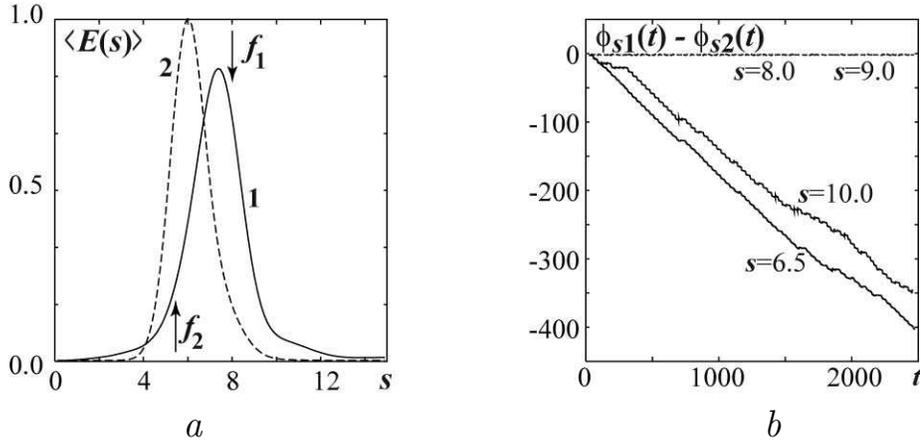}}
\centerline{\large\quad\textit{a}\qquad\qquad\qquad\qquad\qquad\qquad\qquad\quad\textit{b}}
\caption{(\textit{a}) The normalized energy distribution in
wavelet spectrum $\langle E(s)\rangle$ for the first (the solid
line denoted as ``1'') and the second (the dashed line denoted as
``2'') R\"ossler systems. The time scales pointed by means of
arrows correspond to the frequencies $f_1=0.125$ and $f_2=0.154$,
respectively; (\textit{b}) the phase difference
$\phi_{s1}(t)-\phi_{s2}(t)$ for two coupled R\"ossler systems. The
generalized synchronization has been observed \label{fgr:GS2_wvt}}
\end{figure}

Let us consider Fourier spectra of coupled chaotic oscillators
(see Fig.~\ref{fgr:GS2_spectra}). There are two main spectral
components with frequencies $f_1=0.125$ and $f_2=0.154$ in these
spectra. The analysis of behavior of time scales has shown that
both the time scales $s_1=1/f_1=8$ of coupled oscillators
corresponding to the frequency $f_1$ and time scales close to
$s_1$ are synchronized while the time scales $s_2=1/f_2\simeq 6.5$
and close to them don't demonstrate synchronous behavior
(Fig.~\ref{fgr:GS2_wvt},\textit{b}).

The source of such behavior of time scales become clear from the
wavelet power spectra $\langle E(s)\rangle$ of both systems (see
Fig.~\ref{fgr:GS2_wvt},\textit{a}). The time scale $s_1$ of the
drive R\"ossler system is characterized by the large value of
energy while the part of energy falling on this scale of the
response system is quite small. Therefore, the drive system
dictates its own dynamics on the time scale $s_1$ to the response
system. The contrary situation takes place for the time scales
$s_2$ (see Fig.~\ref{fgr:GS2_wvt},\textit{a}). The drive system
can not dictate its dynamics to the response system because the
part of energy falling on this time scale is small in the first
R\"ossler system and enough large in the second one. So, time
scales $s_2$ are not synchronized.

Thus, the generalized synchronization of the unidirectional
coupled R\"ossler systems appears as the time scale synchronized
dynamics, as before another synchronization types do. It is also
clear, why the phase synchronization hasn't been observed in this
case. One can see from the Fig.~\ref{fgr:GS2_spectra} that the
instantaneous phases $\phi_{1,2}(t)$ of chaotic signals
$\mathbf{x}_{1,2}(t)$ introduced by means of traditional
approaches~(\ref{eq:Angle})--(\ref{eq:PoincareSecant}) are
determined by the both frequencies $f_1$ and $f_2$, but only the
spectral components with the frequency $f_1$ are synchronized. So,
the observation of instantaneous phases $\phi_{1,2}(t)$ doesn't
allow to detect the phase synchronization in this case although
the synchronization of time scales takes place.

Thus, one can see that there is a close relationship between
different types of the chaotic oscillator synchronization.
Unfortunately, it is not clear, how one can distinguish the phase
synchronization\footnote{We mean here the phase synchronization
between chaotic oscillators takes place if the instantaneous phase
$\phi(t)$ of chaotic signal may be correctly introduced by means
of (\ref{eq:Angle})--(\ref{eq:PoincareSecant}) and the phase
locking condition~(\ref{eq:PhaseLocking}) is satisfied.} and the
generalized synchronization using only the results obtained from
the analysis of the time scale dynamics and what kind of the
relationship between these synchronization types is. We suppose
that GS and PS are often practically equivalent (when, of course,
it is possible to define correctly the instantaneous phase of
chaotic signal by means of traditional technique). Nevertheless,
this problem should be separately investigated later.

\section{Conclusion}
\label{Sct:Conclusion}

Summarizing this work we would like to note several principal
aspects. Firstly, the traditional approach for the detecting of
the phase synchronization based on the introducing of the
instantaneous phase $\phi(t)$ of chaotic signal is suitable and
correct for such time series which are characterized by the
Fourier spectrum with the single main frequency $f_0$. In this
case the phase $\phi_{s0}$ associated with the time scale $s_0$
corresponding to the main frequency $f_0$ of the Fourier spectrum
coincides approximately with the instantaneous phase $\phi(t)$ of
chaotic signal introduced by means of the traditional approaches
(see also~\cite{Quiroga:2002_Kraskov}). Indeed, as the other
frequencies (the other time scales) don't play a significant part
in the Fourier spectrum, the phase $\phi(t)$ of chaotic signal is
close to the phase $\phi_{s0}(t)$ of the main spectral frequency
$f_0$ (and the main time scale $s_0$, respectively). It is
obvious, that in this case the mean frequencies
$\bar{f}=\langle\dot{\phi}(t)\rangle/2\pi$ and
$\bar{f}_{s0}=\langle\dot{\phi}_{s0}(t)\rangle/2\pi$ should
coincide with each other and with the main frequency $f_0$ of the
Fourier spectrum (see also~\cite{Anishchenko:2004_ChaosSynchro})
\begin{equation}
\bar{f}=\bar{f}_{s0}=f_0.
\end{equation}
If the chaotic time series is characterized by the Fourier
spectrum without the main single frequency (like the spectrum
shown in the Fig.~\ref{fgr:FunnelRoessler},\textit{b}) the
traditional approaches (\ref{eq:Angle})--(\ref{eq:PoincareSecant})
fail. It is clear that one has to consider the dynamics of the
system on all time scales, but it is impossible to do by means of
the instantaneous phase $\phi(t)$ introduced as
(\ref{eq:Angle})--(\ref{eq:PoincareSecant}). On the contrary, our
approach based on the continuous wavelet transform can be used for
both types of chaotic signals.

Secondly, our approach can be easily applied to the experimental
data because it doesn't require any a-priori information about the
considered dynamical systems. Moreover, in several cases the
influence of the noise can be reduced by means of the wavelet
transform (for detail,
see~\cite{alkor:2003_WVTBookEng,Torrence:1998_Wvt,Gusev:2003_wvt}).
We believe that our approach will be useful and effective for the
analysis of physical, biological, physiological etc. data, such
as~\cite{Elson:1998_NeronSynchro,Quiroga:2002_Kraskov,%
Lachaux:2000_WVTSynchro}.

Finally, it is important to note that analysis of the system
dynamics on the different time scales based on the continuous
wavelet transform allows to consider the different types of
behavior of coupled oscillators (such as the complete
synchronization, the lag synchronization, the phase
synchronization, the generalized synchronization and the
nonsynchronized oscillations) from the universal position. It is
clear, that the number of synchronized time scales determines the
type of behavior uniquely. Probably, the quantitative
characteristic of the synchronization measure can be introduced.
This method (with insignificant modifications) can also be applied
to dynamical systems synchronized by the external (e.g., harmonic)
signal.

\section{Acknowledgments}
\label{Sct:Acknowledgments} We express our appreciation to George
A. Okrokvertskhov, Alexander V. Kraskov and Professors
Vadim~S.~Anishchenko and Tatyana E. Vadivasova for valuable
discussions. We thank also Svetlana V. Eremina for the support.

\end{document}